\documentclass[twocolumn,showpacs,preprintnumbers,amsmath,amssymb]{revtex4}
\usepackage{epsfig}
\usepackage{dcolumn}
\usepackage{bm}
\usepackage[usenames]{color}

\begin{document}


\title{Proposal for chiral bosons search at LHC via their unique new signature}

\author{M. V. Chizhov}
\affiliation{Centre for Space Research and Technologies, Faculty of Physics,\\
University of Sofia, 1164 Sofia, Bulgaria}

\author{V. A. Bednyakov}
\affiliation{Dzhelepov Laboratory of Nuclear Problems,\\
\mbox{Joint Institute for Nuclear Research, 141980, Dubna, Russia}}

\author{J. A. Budagov}
\affiliation{Dzhelepov Laboratory of Nuclear Problems,\\
\mbox{Joint Institute for Nuclear Research, 141980, Dubna, Russia}}


\begin{abstract}
    The resonance production of new chiral spin-1 bosons and their detection
    through the Drell--Yan process at the CERN LHC is considered.
    Quantitative evaluations of various differential cross-sections
    of the chiral bosons production are made
    within the CalcHEP package.
    The new neutral chiral bosons can be observed as
    a Breit--Wigner resonance peak
    in the invariant dilepton mass distribution, as usual.
    However, unique new signatures of the chiral bosons exist.
    First, there is no Jacobian peak
    in the lepton transverse momentum distribution.
    Second, the lepton angular distribution
    in the Collins-Soper frame for the high on-peak
    invariant masses of the lepton pairs
    has a peculiar ``swallowtail'' shape.
\end{abstract}


\maketitle

\section{Introduction}

The gauge interactions are the only well-established fundamental
interactions in Nature. Nevertheless, the additional Yukawa
interactions of the Higgs bosons are also necessary for a
self-consistent construction of the Standard Model (SM). They are
still waiting for their experimental verification as a priority part
of the CERN LHC program. Besides this, many other programs and
searches have been approved~\cite{ATLAS,CMS} to explore further the
LHC potential. One of them consists in searching of hypothetical
excited fermions $f^*$ and their magnetic moment (Pauli) type
couplings to ordinary matter
\begin{equation}\label{f*}
    {\cal L}_{\rm excited}^{\color{red}f^*}=
    \frac{g}{2\Lambda}{\color{red}\bar{f}^*}\sigma^{\mu\nu}\!f
    \left(\partial_\mu Z_\nu-\partial_\nu Z_\mu\right)
    +{\rm h.c.},
\end{equation}
where the parameter $\Lambda$ is connected to the compositeness mass
scale of the new physics. Up to now their searches have been
fulfilled at the powerful colliders, such as LEP~\cite{LEP},
HERA~\cite{HERA} and Tevatron~\cite{Tevatron}. Due to their
anomalous type of couplings they lead to a unique experimental
signature for their detection.

In this paper we would like to interpret the interactions (\ref{f*})
from a different point of view, introducing the excited boson states
\begin{equation}\label{Z*}
    {\cal L}_{\rm excited}^{\color{red}Z^*}=
    \frac{g}{2\Lambda}\bar{f}\,\sigma^{\mu\nu}\!f
    \left(\partial_\mu{\color{red} Z^*_\nu}
    -\partial_\nu{\color{red} Z^*_\mu}\right)
\end{equation}
instead of the fermionic ones. It has been shown \cite{TNJL} that
these states with analogous interactions are present among hadron
resonances. Such type of new electroweak heavy bosons $Z^*$ could
also be interesting objects for experimental searches due to their
different couplings to the ordinary fermions in comparison with the
gauge $Z'$ couplings. Alas, this proposal is not popular now and is
not present in the experimental programs, yet. With this note we
would like to establish the {\em status quo\/} of our alternative
proposal for the search for new spin-1 heavy bosons.

In what follows we will concentrate only on resonance production of
the excited neutral heavy bosons at the CERN LHC and their search
via the very clean dilepton Drell--Yan process. To be more concrete
in our predictions we will use a simple phenomenological model of
excited bosons with the chiral couplings as in eq. (\ref{Z*}).

\section{The model}

New heavy neutral gauge bosons are predicted in many extensions of
the Standard Model (SM). They are associated with additional $U(1)'$
gauge symmetries and are generically called $Z'$. The gauge
interactions of these bosons with matter lead to a specific angular
distribution of the outgoing lepton in the dilepton center-of-mass
reference frame with respect to the incident parton
\begin{equation}\label{sV}{\color{blue}
    \frac{{\rm d}\sigma_{Z'}}{{\rm d}\cos\theta^*}\propto 1+
    {\rm ASYM}\cdot\cos\theta^*+\cos^2\theta^*},
\end{equation}
which at present is interpreted as a canonical signature for the
intermediate bosons with spin 1. The coefficient ASYM defines the
backward-forward asymmetry, depending on $P$-parity of $Z'$
couplings to matter.

In addition, another type of spin-1 bosons may exist, which leads to
a different signature in the angular distribution. This follows from
the presence of different types of relativistic spin-1 fermion
chiral currents $\bar{\psi}\gamma^\mu(1\pm\gamma^5)\psi$ and
$\partial_\nu[\bar{\psi}\sigma^{\mu\nu}(1\pm\gamma^5)\psi]$, which
can couple to the corresponding bosons. The hadron physics of the
quark-antiquark meson states provides us with an example of such
kind of interactions and a variety of spin-1 states.

It was pointed out~\cite{TNJL} that {\em three different} quantum
numbers $J^{PC}$ of existing spin-1 mesons, ${\color{blue}1^{--}}$,
${\color{blue}1^{++}}$ and ${\color{red}1^{+-}}$, cannot be assigned
just to {\em two} vector $\bar{q}\gamma^\mu q$ and axial-vector
$\bar{q}\gamma^\mu \gamma^5 q$ quark states. So, the additional
antisymmetric quark tensor currents
$\partial_\nu(\bar{q}\sigma^{\mu\nu}q)$ and
$\partial_\nu(\bar{q}\sigma^{\mu\nu}\gamma^5 q)$ are required, which
also describe vector and axial-vector meson states, but with
different transformation properties with respect to Lorentz group
and with different quantum numbers $1^{--}$ and $1^{+-}$,
respectively. This example demonstrates that both the pure tensor
states, ${\color{red}b_1}$ mesons, and mixed combinations of vector
and tensor states, $\rho$ and $\rho'$ mesons, exist.

The mesons coupled to the tensor quark currents are some types of
``excited'' states as far as the only orbital angular momentum with
$L=1$ contributes to the total angular moment, while the total spin
of the system is zero. This property manifests itself in their
derivative couplings to matter and a different chiral structure of
the interactions in comparison with the gauge ones. In contrast with
the gauge couplings, where either only left-handed or right-handed
fermions participate in the interactions, the tensor currents mix
both left-handed and right-handed fermions. Therefore, like the
Higgs particles the corresponding bosons carry a nonzero chiral
charge. To our knowledge, such bosons were first introduced by
Kemmer~\cite{Kemmer} and they naturally appear in the extended
conformal supergravity theories~\cite{supergravity}.

There are searches for the excited lepton and quark states, but not
for the boson ones. This paper fills this gap considering properties
of the neutral heavy chiral bosons which help us to disentangle
their production at the hadron colliders from other particles. In
order to detect them in the Drell-Yan processes they should couple
to the {\em down\/} type of fermions
\begin{equation}\label{Z*ed}{\color{red}
    {\cal L}_{\rm excited}=\frac{g}{2\sqrt{2}\Lambda}
    \left(\bar{\ell}\sigma^{\mu\nu}\ell+\bar{d}\sigma^{\mu\nu}d\right)
    \left(\partial_\mu Z^*_\nu-\partial_\nu Z^*_\mu\right)}.
\end{equation}
Let us assume for simplicity that $\Lambda$ is equal to the mass of
the new $Z^*$ boson
\begin{equation}\label{M}
     M \approx 1~{\rm TeV}
\end{equation}
and $g$ being the coupling constant of the $SU(2)_W$ weak gauge
group. For comparison we will consider topologically analogous gauge
interactions of the $Z'$ boson
\begin{equation}\label{Z'ed}{\color{blue}
    {\cal L}_{\rm gauge}=\frac{g}{2}
    \left(\bar{\ell}\gamma^{\mu}\ell+\bar{d}\gamma^{\mu}d\right)
    Z'_\mu}
\end{equation}
with the same mass $M$. The coupling constants are chosen in such a
way that all fermionic decay widths in the Born approximation of the
both bosons are identical. It means that their total production
cross-sections at the hadron colliders are nearly equal up to
next-to-leading order corrections. Their leptonic decay width
\begin{equation}\label{Gl}
    \Gamma_\ell=\frac{g^2}{48\pi}M\approx 2.8~{\rm GeV}.
\end{equation}
is sufficiently narrow so that they can be identified as resonances
at the hadron colliders in the Drell--Yan process.

\section{The experimental signature}

\begin{figure}[th]
\epsfig{file=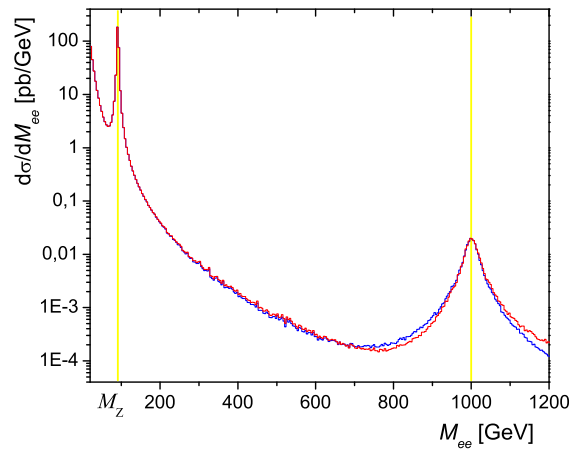,width=8.5cm} \caption{\label{fig:1} The
invariant dilepton mass distributions for the gauge $Z'$ boson
(blue) and the excited chiral $Z^*$ boson (red) with the Drell--Yan
SM background at the CERN LHC.}
\end{figure}

Up to now, the excess in the Drell--Yan process with high-energy
invariant mass of the lepton pairs remains the clearest indication
of the heavy boson production at the hadron colliders. So, using
only a modest integrated luminosity of 200 pb$^{-1}$ collected
during RUN~II, the D0 Collaboration puts tight restrictions on the
$Z'$ masses for the different models from the dielectron
events~\cite{D0_4375}: $M_{Z'_{SM}} < 780$~GeV, $M_{Z'_\eta} <
680$~GeV, $M_{Z'_\psi} < 650$~GeV, $M_{Z'_\chi} < 640$~GeV and
$M_{Z'_I} < 575$~GeV. Comparable statistics in the dimuon channel
leads approximately to the same constraint $M_{Z'_{SM}} <
680$~GeV~\cite{D0_4577}. The CDF constraints from the dielectron
channel are based on more data, 1.3 fb$^{-1}$, which leads to
tighter restrictions~\cite{CDF_8694}: $M_{Z'_{SM}} < 923$~GeV,
$M_{Z'_\eta} < 891$~GeV, $M_{Z'_\psi} < 822$~GeV, $M_{Z'_\chi} <
822$~GeV and $M_{Z'_I} < 729$~GeV. Therefore, our mass choice
(\ref{M}) for the new bosons does not contradict the Tevatron
constraints. At the same time such bosons could be observed as
resonance peaks on the $Z$ boson tail in the invariant dilepton mass
distribution at the LHC (Fig.~\ref{fig:1} \footnote{Here and in the
following the CalcHEP~\cite{CalcHEP} package will be used for the
numeric calculations of various distributions with a CTEQ6M choice
for the proton parton distribution set.}) already in the first days
of the physical runs.

\begin{figure}[th]
\epsfig{file=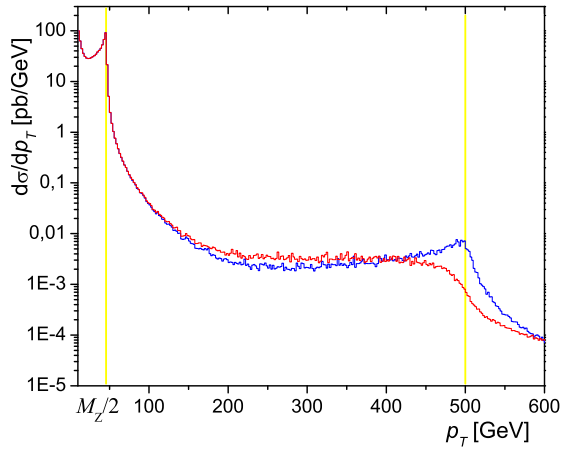,width=8.5cm} \caption{\label{fig:2} The
differential cross-sections for the gauge $Z'$ boson (blue) and the
excited chiral $Z^*$ boson (red) with the Drell--Yan SM background
as functions of the lepton transverse momentum at the CERN LHC.}
\end{figure}

The peaks in the invariant mass distributions originate from the
Breit--Wigner propagator form, which is the same both for the gauge
and chiral bosons in the Born approximation. However, the common
wisdom, that a peak in the invariant mass distribution of the two
final particles must correspond to the Jacobian peaks in their
transverse momentum distributions $p_T$, is not valid for the chiral
bosons due to the following fact. The main feature of the
interactions (\ref{Z*ed}) consists in different angular distribution
of final fermions~\cite{two}
\begin{equation}\label{sT}{\color{red}
    \frac{{\rm d}\sigma_{Z^*}}{{\rm d}
    \cos\theta^*}\propto\cos^2\theta^*}
\end{equation}
in comparison with the distribution (\ref{sV}) for the gauge
interactions. It leads to a stepwise lepton transverse momentum
distribution, rather than to the Jacobian peak at the kinematical
endpoint $M/2$ for the gauge bosons (Fig.~\ref{fig:2}).
Therefore, already the lepton transverse momentum distribution
demonstrates a difference between the gauge and chiral bosons. In
order to make more substantial conclusions, let us investigate other
distributions selecting only ``on-peak'' events with the invariant
dilepton masses in the range 800~GeV~$<M_{\ell\ell}<$~1200~GeV.

\begin{figure}[th]
\epsfig{file=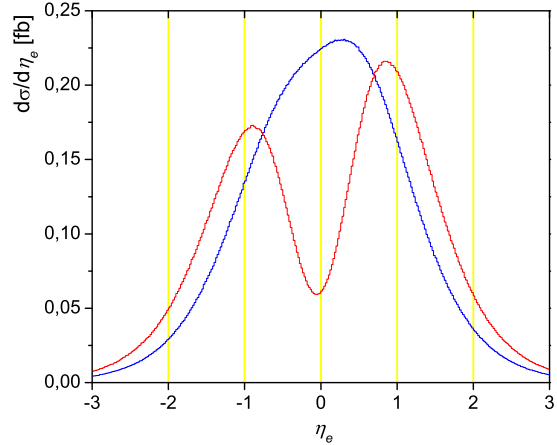,width=8.5cm} \caption{\label{fig:3} The
differential cross-sections for the gauge $Z'$ boson (blue) and the
excited chiral $Z^*$ boson (red) decaying to a lepton pair with the
invariant mass 800~GeV~$<M_{\ell\ell}<$~1200~GeV as functions of the
lepton pseudorapidity at the Fermilab Tevatron.}
\end{figure}

\begin{figure}[th]
\epsfig{file=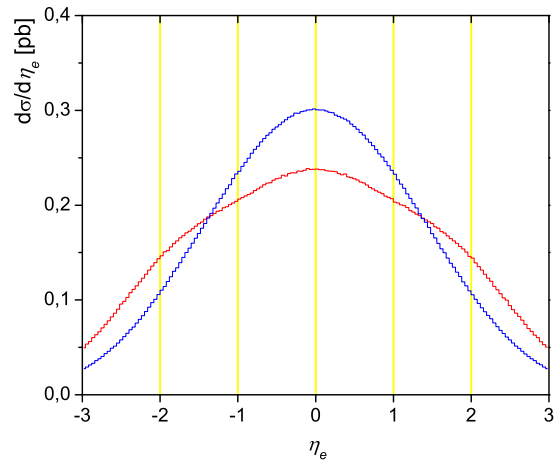,width=8.5cm} \caption{\label{fig:4} The
differential cross-sections for the gauge $Z'$ boson (blue) and the
excited chiral $Z^*$ boson (red) decaying to a lepton pair with the
invariant mass 800~GeV~$<M_{\ell\ell}<$~1200~GeV as functions of the
lepton pseudorapidity at the CERN LHC.}
\end{figure}

According to the eq. (\ref{sT}), there exists a characteristic
plane, perpendicular to the beam axis in the parton rest frame,
where the emission of the final-state pairs is forbidden. The
nonzero probability in the perpendicular direction in the laboratory
frame is due to the longitudinal boosts of colliding partons. So, at
the Fermilab Tevatron the production of such heavy bosons occurs
almost at the threshold with approximately zero longitudinal
momenta. Hence, the lepton pseudorapidity distribution for the
chiral bosons has a minimum at $\eta_\ell=0$ (Fig.~\ref{fig:3}).
On the other hand the CERN LHC is sufficiently powerful to produce
heavy bosons with a mass $M=1$~TeV with high longitudinal boosts.
Therefore, the pseudorapidity distributions for the gauge and chiral
bosons at the LHC look similar (Fig.~\ref{fig:4}).

\begin{figure}[th]
\epsfig{file=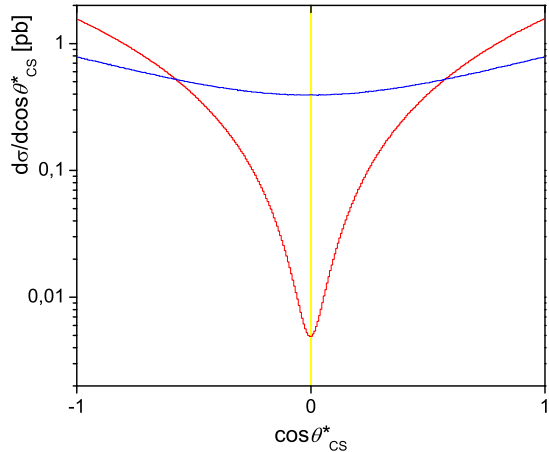,width=8.5cm} \caption{\label{fig:5} The
differential cross-sections for the gauge $Z'$ boson (blue) and the
excited chiral $Z^*$ boson (red) decaying to a lepton pair with the
invariant mass 800~GeV~$<M_{\ell\ell}<$~1200~GeV as functions of
$\cos\theta^*_{\rm CS}$ at the CERN LHC.}
\end{figure}

Crucial confirmation of the existence of the new interactions
(\ref{Z*ed}) should come from the analysis of the angular
distribution of the final leptons with respect to the boost
direction of the heavy boson in the rest frame of the latter (the
Collins-Soper frame). In the Fig.~\ref{fig:5} we compare the
differential cross-sections for the gauge $Z'$ boson and the excited
chiral $Z^*$ boson decaying to the lepton pairs with the invariant
mass 800~GeV~$<M_{\ell\ell}<$~1200~GeV as functions of
$\cos\theta^*_{\rm CS}$.
Instead of a smoother angular distribution for the gauge
interactions, a peculiar ``swallowtail'' shape of the chiral boson
distribution occurs with a dip at $\cos\theta^*_{\rm CS}=0$. It will
indicate the presence of the new interactions. Neither scalars nor
other particles possess such a type of angular behavior.

\section{Conclusions}

In this paper we have considered the experimental signatures of the
excited chiral heavy bosons $Z^*$ and compared them with the gauge
$Z'$ bosons. It has been stressed that the chiral bosons have a new
angular distribution, yet unknown for experimentalists. It leads to
an absence of the Jacobian peak in the transverse momentum
distribution and to a profound dip in the angular distribution at
the rest frame of the heavy chiral boson. These features could help
to discriminate the chiral boson production from a resonance
production of other particles at the hadron colliders.

\section*{Acknowledgements}
We are grateful to V. G. Kadyshevsky, M. D. Mateev, N. A.
Russakovich, I. R. Boyko and R. V. Tsenov for support and fruitful
cooperation.


\end{document}